\documentclass[pre,showpacs,floatfix,twocolumn]{revtex4}
\usepackage{graphicx}
\usepackage{amssymb}
\usepackage{dcolumn}
\usepackage{bm}
\begin{document}

\title{Morphologies of expansion ridges of elastic thin films onto a substrate}
\author{E. A. Jagla}
\affiliation{Centro At\'omico Bariloche, Comisi\'on Nacional de Energ\'{\i}a At\'omica, 
(8400) Bariloche, Argentina}

\begin{abstract}

We consider a model of a thin film elastically attached to a 
rigid substrate. 
In the case in which the film expands relative to 
the substrate and assuming certain non-linear elastic behavior of the film, 
expansion ridges may appear,
in which the material has collapsed, and the density is higher. By 
studying numerically this process, the possible morphologies of these 
collapsed regions are presented. They range from circular spots and 
straight stripes, to wiggle polygonal patterns and ring-shaped
domains. The similarity of some of these results with patterns observed in 
delamination of thin films and bi-phase epitaxial growth is emphasized.

\end{abstract}
\maketitle

\section{Introduction}

When the surface of a material 
contracts with respect to the underlying part, 
tensile stresses appear in it that can lead
to the formation of a crack pattern onto the surface\cite{barro}.
Mud cracking can be considered to be the 
prototype of this surface fragmentation process. 
The main ingredients of surface fragmentation are a (quasi) two-dimensional 
film attached to a substrate (we always assume here the substrate is rigid)
and a greater expansion of the substrate compared to the film as a function of
some external control variable (typically humidity concentration or temperature).
The film and substrate can be the same material, as in mud cracks, and 
in that case it is only a difference in humidity concentration or temperature what
identifies film and substrate.

We will concentrate here in a process that in a certain sense is the inverse 
of surface fragmentation: we consider the film expanding with respect to the 
substrate. Due to the coupling to the substrate, the greater expansion of 
the film generates compressive stresses into it. 
No important effects are expected if the film responds linear 
elastically to any deformation. But if 
the film can collapse upon compression, a coexistence of collapsed and 
non-collapsed regions in the film is expected. The presentation and 
discussion of the different morphologies of these collapsed regions are the 
main aim of this paper.

The two main features that fix the morphology of collapsed regions are the degree of mismatch between 
film and substrate, and the kind of non-linear elastic behavior of the film, in particular, the characteristics 
of the collapsed state. Different possibilities are studied here.
Most of the results to be presented correspond to the case in which the system is 
isotropic in the plane of the film (the $x$-$y$ plane), but some results for a model 
with square symmetry will also be presented. In the next Section we present the model and details on the
simulation technique. In Section III we present the main results, and in Section IV we discuss 
two experimental situations in which the present model can be applied, namely,
delamination patterns of thin film and structures appearing during bi-phase epitaxial growth.
Section V contains some summary and conclusions.

\section{The model}

The model to be used is an extension of that used in \cite{marconijagla,jagla1} to describe fracture (see also
\cite{carta} and \cite{shenoy}). It considers the film in a two-dimensional 
approximation as described by the horizontal displacement field ${\bf u}({\bf r})$.
From these variables 
the two dimensional strain tensor is calculated as
$\varepsilon_{ij}\equiv 1/2(\partial u_i/\partial x_j+\partial u_j/\partial x_i)$.
For convenience, instead of $\varepsilon_{ij}$ we will use the following variables

\begin{eqnarray}
e_1&\equiv&(\varepsilon_{11}+\varepsilon_{22})/2\nonumber\\
e_2&\equiv&(\varepsilon_{11}-\varepsilon_{22})/2\\
\label{dos}
e_3&\equiv&\varepsilon_{12}=\varepsilon_{21}\nonumber
\end{eqnarray}
These three variables 
are not independent. They satisfy the St. Venant compatibility constraint\cite{carta,shenoy,chandra}, 
which reads

\begin{equation}
(\partial^2_x+\partial^2_y)e_1-(\partial^2_x-\partial^2_y)e_2-2\partial_x\partial_y e_3=0.
\end{equation}

The free energy density of the system contains three terms: a local 
term $f_0$, a gradient term $f_{\nabla}$, and a substrate interaction term 
$f_{subs}$. The existence of a collapse transition for the film is encoded 
in the form of the local free energy term $f_0$. If we intend to 
describe an isotropic material, only rotationally invariant combinations 
of the basic variables $e_i$ should enter the free energy. 
Those that can be constructed from the $e_i$ are $e_1$ itself, and 
$e_d \equiv \sqrt{e_2^2+e_3^2}$.
A perfectly elastic material is described by a local free energy having a 
quadratic minimum, namely its local free energy is of the form
\begin{equation}
f_0^{elastic}=B(e_1-e_1^0)^2+\mu(e_2^2+e_3^2)
\end{equation}
where $B$ and $\mu$ are respectively proportional to the bulk and shear modulus of the material. 
With the present choice, in a completely relaxed state the system has $e_1=e_1^0$, $e_2=e_3=0$.

\begin{figure}
\includegraphics[width=8cm,clip=true]{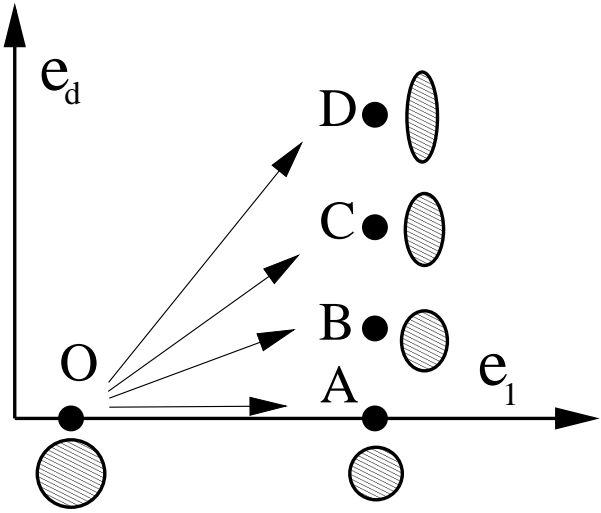}
\caption{Different possibilities for the location of the collapsed minimum in the
$e_1$-$e_d$ plane. The elastic minimum is located at O. The sketches 
show qualitative an
originally circular piece of
material at O and its collapsed form for different locations of the collapsed minimum. 
In cases {\bf A} and {\bf B} the an original circular piece of material 
contracts in all directions, whereas for {\bf C} and {\bf D} there is a
direction in which it actually expands (chosen here to be the vertical direction).}
\label{f1}
\end{figure}

If the material can collapse, $f_0$ must have some other minimum at some $e_1^C$, $e_d^C$ describing 
the collapsed state. The position of the collapsed minimum (in particular, the ratio
$\nu\equiv e_d^C/(e_1^C-e_1^0)$) will have important 
consequences in the morphologies of the collapsed regions that will be observed. 
Different locations of the collapsed minimum will be explored and discussed here. They are 
qualitatively depicted in Fig. \ref{f1} as cases {\bf A}, {\bf B}, {\bf C}, and  {\bf D}.
In most cases the morphologies of the collapsed patterns depend only on the position of the collapsed 
minimum with respect to the elastic minimum, the detailed form of $f_0$ being of minor importance.
Only in case {\bf A}, there are important variations depending on the
value of the shear modulus of the collapsed state $\mu^C$ with 
respect to the value in the normal phase $\mu$.
Thus,
two cases will be distinguished: {\bf A1}, in which  $\mu^C$ is greater 
or equal $\mu$, and {\bf A2}, in which $\mu^C$ is lower than $\mu$.  
Note that when the collapsed state has $e_d^C\ne 0$, it corresponds actually to a ring of minima in the $e_1$, 
$e_2$, $e_3$ space. In other words, the collapse of a circular piece of material to a state with $e_d\ne 0$
produces an elongated object, but the orientation of this object in the $x$-$y$ plane 
can be any (in the drawings of Fig. \ref{f1} this
direction was chosen as the vertical one).
The explicit form of the $f_0$ part of the free energy for the cases {\bf A}-to-{\bf D}
is given in the appendix.

Gradient terms in the free energy will be taken in the form
\begin{equation}
f_{\nabla}=\sum_{i=1,2,3} \alpha_i (\nabla e_i)^2,
\label{grad}
\end{equation}
where 
$\alpha_2=\alpha_3$ should be chosen to retain rotational invariance. In all results below 
we take all $\alpha_i\equiv \alpha$.

The elastic interaction with the substrate is easily written in terms of the displacement variables 
${\bf u}$:
\begin{equation}
 f_{subs}= \frac{\gamma}{2} |{\bf u}({\bf r})|^2
\end{equation}
where $\gamma$ measures the stiffness of this interaction.
As we take the components of $\varepsilon$ to be our basic variables, we have to recast this energy in terms
of them. This is more easily done in the Fourier space, and the result can be easily written 
after integration over the whole system as
\begin{equation}
\int d{\bf r}^2 f_{subs}=2\gamma\int' d{\bf k}^2 \frac{|\tilde e_2({\bf k})|^2+|\tilde e_3({\bf k})|^2}{k^2},
\label{subst}
\end{equation}
Where $\tilde e_i({\bf k})$ are the Fourier transforms of the original $e_i({\bf r})$. 
Here the tilde in the integral indicates that the $k=0$ mode is excluded. To avoid a divergent energy
contribution, the value of
$\tilde e_1({\bf k}=0)=\bar e_1$ (where the bar notes a spatial average) has to be adjusted for the system 
to fit on average to the substrate, namely,
the mismatch with the substrate is incorporated in the model precisely through the value given to $\bar e_1$.

The existence of non trivial spatial patterns in the model appears as a consequence of a competition between gradient and
substrate term. In fact, in Fourier space a mode of wave vector $\sim k$ and amplitude 1 produces a contribution to the
energy of the order of $\alpha k^2$ from the gradient terms, and $\gamma/k^2$ from the substrate term. The sum of this two
contribution has a minimum at a value of $k$ of the order of $(\gamma/\alpha)^{1/4}$. This is the order of magnitude of the
main spatial variations that will be seen to appear in the simulations.

The equations of motion are taken to be of the overdamped form, namely
\begin{equation}
\frac{\partial e_i}{\partial t}=-\lambda \frac{\delta F}{\delta e_i}~~~~~(i=1,2,3)
\label{eqs}
\end{equation}
where
\begin{equation}
F=\int d{\bf r}^2 \left (f_0+f_{\nabla}+f_{subs}\right ).
\label{f}
\end{equation}
The Saint Venant constraint is implemented by the use of Lagrange multipliers.
Full details can be seen in \cite{marconijagla} and \cite{shenoy}.

\section{Results}

For each set of parameters, the system will settle down in a configuration that minimizes the total energy of the system.
We will see that in general, metastable states appear very often in the simulations.
In order to get as 
close as possible to the true ground state configuration, an annealing process 
was implemented in which a stochastic term was added to the right hand side of
Eq. (\ref{eqs}), and the intensity of this term was progressively reduced down 
to zero during the simulation. 
The final configurations obtained are good examples of the
typical morphologies favored by the competition of the different energy terms.

Results will be presented for a unique set of values of the coefficients $\alpha$ and $\gamma$ in the gradient
and substrate interaction terms [Eqs. (\ref{grad}) and (\ref{subst})]. 
It can be shown that a change of these parameters can be
absorbed in a rescaling of the spatial coordinate and a global redefinition of
the free energy\cite{jagla}. We have chosen the values $\alpha=3$, $\gamma=0.01$, 
for which the expected spatial scale of the structures to be seen (based on the estimation in the previous section) 
is of the order of ten 
mesh parameters. Thus, the geometrical features of the patterns will
be reasonably larger than the discretization of the numerical mesh, and this is
appropriate to eliminate spurious effects associated to this discreteness.

The configurations obtained are presented in Figs. \ref{a1} to \ref{d}, corresponding 
to cases {\bf A}-to-{\bf D}, respectively. Each panel in each figure represents a 
different value of $\bar e_1$, i.e., 
a different degree of mismatch with the substrate. 
Each main plot shows the spatial distribution of $e_1$ or $e_d$ (depending on which is more representative
in each case).
The scale for this plot is indicated in the bar of the inset. 
The inset also shows the combined distribution of $e_d$ vs. $e_1$ for all elements in the system. 
We now discuss separately each case. 

Except for the
existence of the substrate, case {\bf A} is qualitatively analogous to a well studied case, which is applicable for
instance to phase separation in alloys\cite{sagui,onuki}. In the case in which 
the collapsed state has the same shear modulus than the original state 
(case {\bf A1}), there is strong evidence that the possible ground states correspond to 
a striped phase, a bubble phase, 
or a uniform phase depending on the value of $\bar e_1$.
The patterns obtained in the numerical simulations presented here (Fig. \ref{a1}) are in reasonable agreement with these
results, although it is clear that the perfect striped phase is not easily obtained. The reason for this
discrepancy is that there is a large gain in entropy when the stripes 
disorder, and it is very difficult numerically
to get rid of this effect. 

In the case in which the collapsed phase has a lower shear modulus than the
original phase (case {\bf A2}, Fig. \ref{a2}), bubbles of the softer phase tend to be unstable with respect to
elongation\cite{sagui,onuki}. This is obvious in Fig. \ref{a2} compared to Fig. \ref{a1}. Note that the effect is not symmetric:
if the mismatch is such that the minority phase corresponds to the more rigid phase, then the bubbles
remain stable. In both cases {\bf A1} and {\bf A2}, for even larger mismatch (not shown)
all the system becomes uniformly collapsed, with $e_1({\bf r})=\bar e_1$, $e_2({\bf r})=e_3({\bf r})=0$.

For case {\bf B} (Fig. \ref{b}) we observe a stronger tendency to form stripes and 
polygonal patterns. 
There is also an important difference with case 
{\bf A2} for large mismatch (lower-right panel):
Now each element prefers to be as close as possible to the collapsed 
minimum, that locates at a non-zero value of $e_d$.
However, not all elements can have the same values of 
$e_2$ and $e_3$, since the spatial averages $\bar e_2$ 
and $\bar e_3$
should be zero. The existence of the substrate also 
discourages uniform phases with constant $e_2$ or $e_3$. 
The configuration of the system is such that individual 
elements tend to be distributed close to the ring 
of minima in the $e_2$-$e_3$ plane. In real space, 
singularities appear at which $e_2=e_3=0$, and around them
the configuration point in the $e_2$-$e_3$ plane rotates 
$2\pi$, clockwise, or counterclockwise, thus defining
`vortex-like' or `anti-vortex-like' defects. Density 
of these defects is mainly controlled by the strength of the 
interaction to the substrate (density tends to zero for 
vanishing interaction).

\begin{figure}
\includegraphics[width=8cm,clip=true]{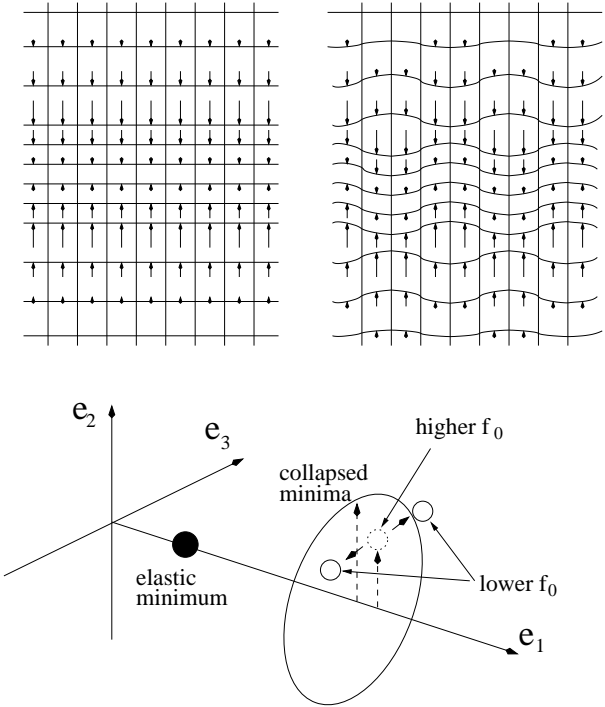}
\caption{(a) The structure of a straight collapsed region. Arrows indicate qualitatively the displacement ${\bf u}({\bf r})$.
This configuration can be unstable with respect to the deformation shown in (b), for certain forms of the local free energy
$f_0$.
In fact, part (c) illustrates how a piece of material collapsed in a single direction (represented by the dotted circle)
can lower its energy by generating a finite shear distortion $e_3$
if this brings it closer to the ring of minima of the potential.}
\label{osci}
\end{figure}

Cases {\bf C} and {\bf D} (Figs. \ref{c} and \ref{d}) present an interesting 
difference in morphology with respect to previous cases:
Stripes of collapsed regions present now a wavy structure.
In some cases, we observe even the existence or 
ring shaped collapsed regions. The origin of 
wiggling collapsed stripes is of course dictated by the 
tendency of the model to minimize the energy. An 
elemental demonstration of the instability of
straight collapsed regions with respect to wiggled 
patterns will be presented now.
Consider the sketch of Fig. \ref{osci}(a) of a straight 
collapsed region along the $x$ direction. The arrows in the plot are a
schematic representation of the displacement field ${\bf u}({\bf r})$. Suppose 
that we have determined the best form of
the function ${\bf u}({\bf r})=u_0(y)\hat y$ in order to minimize 
the energy. We will see that in some circumstances the energy
can be reduced further by changing of ${\bf u}({\bf r})=u_0(y)\hat y$ 
to a new displacement field of the form 
${\bf u}({\bf r})=u_0(y-A\sin(kx))\hat y$. Note that this 
represents a wiggling pattern with oscillation amplitude $A$ and wave
vector $k$ (as sketched in Fig. \ref{osci}(b)). 
By calculating the form of our fundamental variables 
$e_1$, $e_2$ and $e_3$, and
plugging these values into the
different term of our total energy, we see that the 
substrate energy does not change at all. The
gradient term is increased in a quantity (per unit length 
in the $x$ direction) of the order of $A^2k^2d$, where $d$ is the
thickness of the collapsed region. The local term changes 
only because of the existence of a non-zero value of $e_3$. This change is
also proportional to $A^2k^2d$, and the overall factor can be 
negative (i.e., we can have an energy gain) if $\nu$ 
($=e_d^C/(e_1^C-e_1^0)$) is sufficiently large, as depicted in Fig.
\ref{osci}(c). Then we see that if this quadratic decrease of 
energy overcomes the quadratic increase due to the gradient
terms, a wavy stripe is expected. This lowest order sketchy argument 
does not provide values for the amplitude $A$ and wave vector $k$ of 
the distortion. They could in
principle be obtained by considering  higher order terms 
in the expansions. We failed in doing such a calculation. We only
notice that from the results of the numerical simulations 
(Figs. \ref{c} and \ref{d}) it seems plausible that both $A$ and the wave length $k^{-1}$
are of the order of the stripe thickness $d$.

\section{Practical realizations}

We have analyzed an idealized situation of a material that has two well defined elastic 
configurations of minimum energy. We have shown that non-trivial morphologies may appear when a 
quasi-two-dimensional piece of such a material is uniformly attached to a rigid substrate.
In searching for practical realizations, it will be difficult to find that all the above assumptions
are satisfied, however, our model can be a good idealized case to study 
some practical problems. Two examples will be presented now.

The first case corresponds to the patterns observed in delamination of thin films\cite{delam}.
In this case, typically, a film is grown onto a substrate, and due to
chemical differences between substrate and film (which may be enhanced 
by temperature changes) elastic stresses develop between them.
If the stresses are compressive into the film, this may induce the buckling up
of part of the film, giving rise to patterns that have been 
termed `telephone-cord' like, due to their wavy appearance. In some cases, wavy polygonal patterns of
buckled regions are observed, remarkably similar to those in our Figs. \ref{c} and \ref{d}. In delamination patterns
the wave length and amplitude of the undulations observed are both of the order of the thickness of the buckled
region.

\begin{figure}
\includegraphics[width=8cm,clip=true]{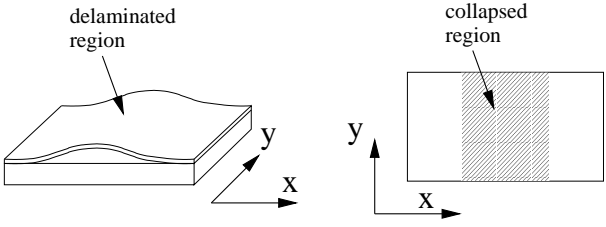}
\caption{A stripe of delaminated thin film, and the qualitative equivalent as a region of collapsed material in our model.}
\label{bck}
\end{figure}

This delamination process has some features that are {\em a priori} very different from the assumptions in our 
model. Two things look in fact very different: in delamination, the three dimensional nature of the process
plays an important role. In contrast,
our model considers a two dimensional geometry. Secondly, in delamination the interaction to the substrate is lost 
in the buckled regions, whereas in our model
this interaction is always present. However, the similar appearance
of some of our results with the delamination phenomenology is very suggestive. We think that a justification 
of the similarity can be based on the following considerations: a stripe of buckled material
is qualitatively depicted in Fig. \ref{bck}(a). If we want to make an effective two-dimensional representation 
of this buckled film, we should associate to the buckled region a piece of the two dimensional model
with higher density. Note also that the orientation of the buckled stripe can be any in the $x$-$y$ plane, but that 
the contraction that induces the density increase occurs only perpendicularly to the buckled ridge. This means 
that we can associate buckling with our collapse transition with $\nu\simeq 1$. In delamination, the
possibility of buckling depends on the possibility of detaching from the substrate. Although in our model
the interaction with the substrate is always present, the free energy with two minima takes phenomenologically into 
account the two configurations of the film: attached and unbuckled (represented by the isotropic minimum 
of our model) and detached and buckled (represented by the finite $e_d^C$ collapsed state).

A second problem to which our model can be applied is the epitaxial growth of competing phases with different
crystallographic parameters\cite{bifase}. In fact, if the growing is coherent (to some extent), elastic
stresses will be accumulated during growth, and which phase is chosen locally to grow the material 
will be dictated by a tendency  to minimize the total elastic energy of the system. 
Thus, our model with two minima must be thought as describing the two different phases that can be 
chosen in the growth process.
But in this case
one of the previously used assumptions needs to be changed: epitaxy is strongly dependent on the
crystalline structures of the phases, and isotropy of the model in the $x$-$y$ plane is not justified. 
However, the model can be easily modified
to account for anisotropies in the $x$-$y$ plane. We will present here only a single example of the possible outcome
obtained in an anisotropic model. Let us consider that the substrate has a square symmetry (assumed to be aligned with the
$x$-$y$ coordinate axis), that one of the growing phases 
has also a square symmetry with a larger unit cell volume, and that the second growing phase has a rectangular symmetry with a
lower unit cell volume than that of the substrate. This situation corresponds to a local free energy $f_0$ having an elastic
minimum as usual, and a collapsed minimum localized now at $e_2=\pm e_2^C$, $e_3=0$. In this way the ring of minima in the
$e_2$-$e_3$ plane of the previous cases is replaced by a couple of minima. The exact form of 
$f_0$ we use is given in the Appendix.
The morphologies obtained are shown in Fig. \ref{rect}. As
expected, the existence of a rectangular phase reflects clearly on the morphology of the patterns.
Note how upon change of the fraction of rectangular phase (which is dictated, as usual, by the degree of mismatch with the
substrate), the morphology changes from stripes of rectangular phase parallel
to the axis (for low density of rectangular phase), to stripes of rectangular phase along the diagonals (for the case of large
fraction of rectangular phase).
It is worth to be noticed that the present case can also be interpreted in the context of a 
model recently introduced to describe a square-to-rectangular martensitic transformation\cite{carta,shenoy}. 
In fact, we can consider the isotropic minimum as representing the austenite phase, and the collapsed minima as
the martensite phase, in its two different variants: $e_2=\pm e_2^C$.
In the case of Refs. \cite{carta,shenoy} (in which a substrate has not been included), 
volume change during martensitic transformation is
assumed to be negligible, and a phase of diagonal stripes similar 
to that in the lower plots of  Fig. \ref{rect} has been observed
\cite{carta,shenoy} (note that in our case the width of the 
diagonal stripes is fixed by the strength of the interaction with the
substrate). From the results of the first plots of Fig. \ref{rect} we 
can conclude that a phase of stripes along the $x$ and
$y$ directions is favored instead when the transformation has an 
appreciable volume change and a finite fraction of the
austenite phase remains in the system.

\section{Summary and Conclusions}

Expansion ridges are analog to surface cracks, with the difference that they appear
when a film expands with respect to the substrate (contrary to cracks, 
that appear when the film contracts with respect to the underlying material).
A non-linear elastic behavior of the film, 
namely the possibility of a collapse upon applied stress is necessary 
for expansion ridges to appear.
We have simulated expansion ridges through an elastic model that uses the components of the strain tensor
as fundamental variables. We have shown how complex morphologies can appear due to competitive elastic interactions in the
system. By changing the parameters of the model, collapsed regions in the form of bubbles, straight stripes, undulating
stripes and rings have been obtained. We have argued that the model can be applied to
understand the characteristics of delamination patterns in thin films, and bi-phase epitaxial growth.
We have concentrated in the description of the morphologies, and then our presentation has been qualitative to a large
extent. A few important things can be highlighted: The relevant spatial scale of the patterns has been seen to be fixed by a competition between interaction with the
substrate and gradient effects. In the case in which the collapse is isotropic, we have qualitatively reproduced 
well known results in the field of alloy decomposition.
Wiggle stripes of collapsed material appear when the collapse of an elemental circular piece
of material is such that contraction in one direction is accompanied by expansion in the perpendicular direction. We have
compared these wiggling patterns with those observed in delamination of thin films. We have also provided an example showing
that anisotropy can be easily incorporated in the model, and showed that the anisotropic case may have relevance
for the study of bi-phase epitaxial growth. More detailed studies an comparisons on each of these individual realizations is
in progress.

\begin{figure}
\includegraphics[width=8cm,clip=true]{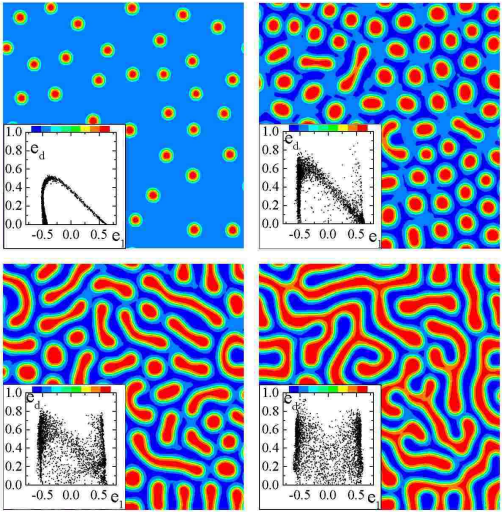}
\caption{(Color online) Spatial configurations obtained for case {\bf A1}, for values of $\bar e_1$ equal to $-0.4$, $-0.2$, $-0.1$, and 0.2
(from left to right and from top to bottom). The shadows indicate values of $e_1$, with the
scale indicated in the inset. In the inset plots, the distribution of $e_d$ vs. $e_1$ is shown for all elements of the
system.}
\label{a1}
\end{figure}

\begin{figure}
\includegraphics[width=8cm,clip=true]{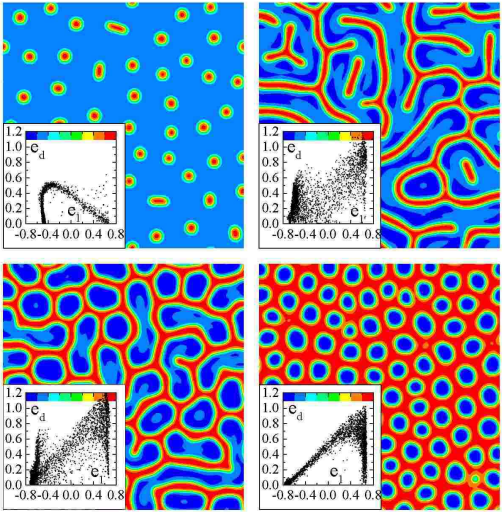}
\caption{(Color online) Same as Fig. \ref{a1} for case {\bf A2}, and values of $\bar e_1$ of $-0.4$, $-0.2$, $-0.1$, and $0$.}
\label{a2}
\end{figure}

\begin{figure}
\includegraphics[width=8cm,clip=true]{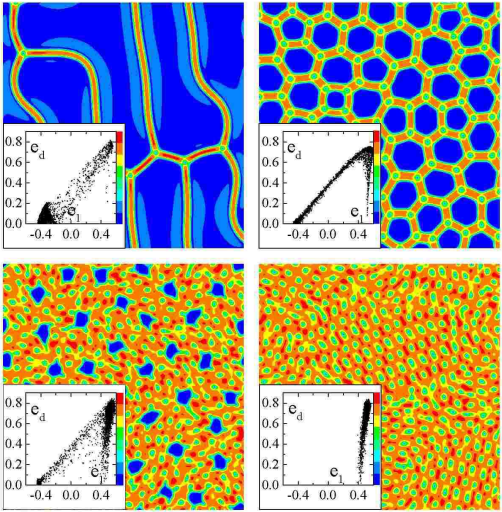}
\caption{(Color online) Same as Fig. \ref{a1} for case {\bf B}, and values of $\bar e_1$ of $-0.2$, $0$, $0.4$, and $0.5$ (note that in this
and next two Figures, the main plot shows values of $e_d$, instead of $e_1$).}
\label{b}
\end{figure}

\begin{figure}
\includegraphics[width=8cm,clip=true]{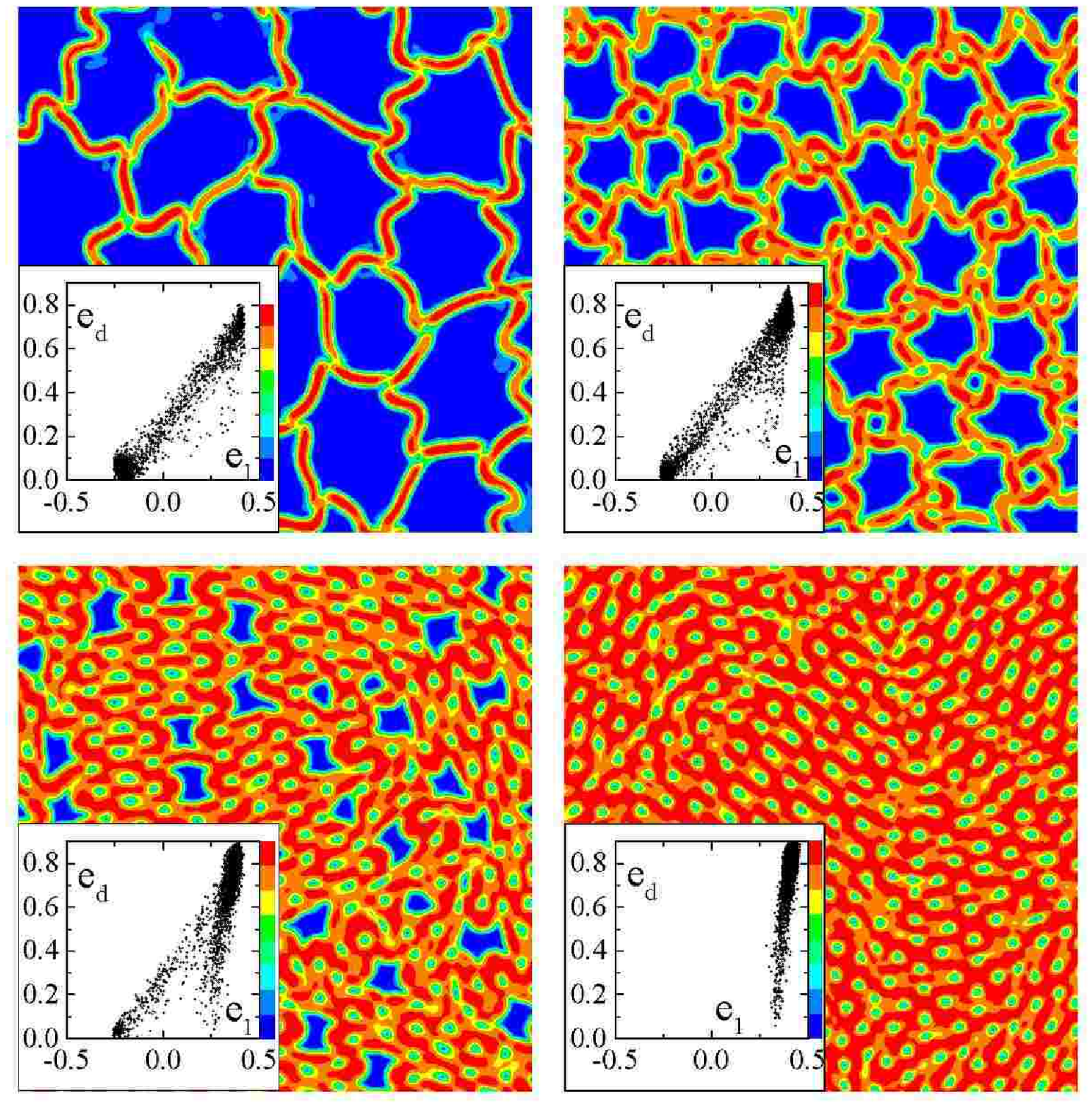}
\caption{(Color online) Same as Fig. \ref{b} for case {\bf C}, and values of $\bar e_1$ of $-0.05$, $0.1$, $0.3$, and $0.4$.}
\label{c}
\end{figure}

\begin{figure}
\includegraphics[width=8cm,clip=true]{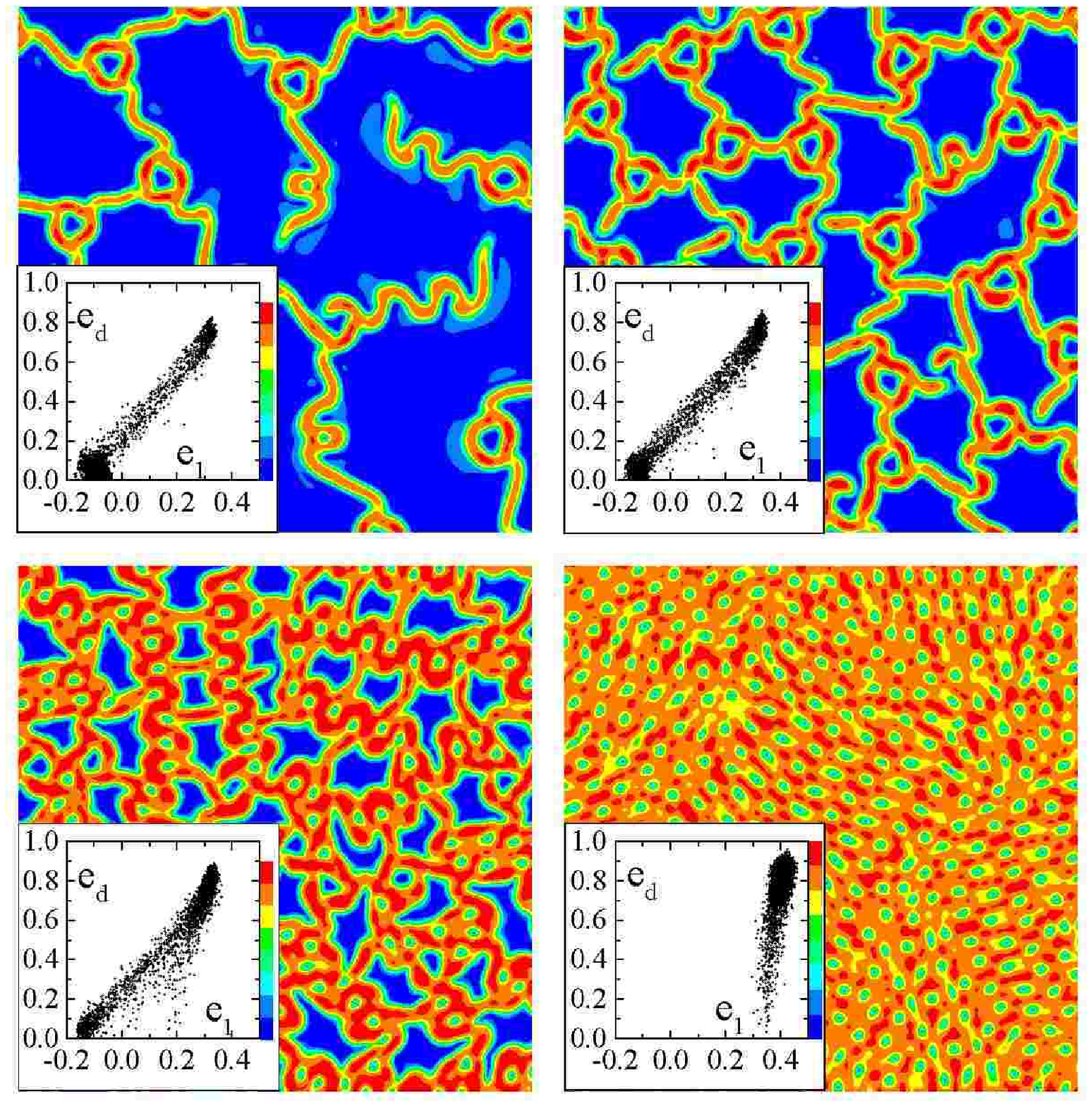}
\caption{(Color online) Same as Fig. \ref{b} for case {\bf D}, and values of $\bar e_1$ of $-0.005$, $0.05$, $0.2$, and $0.4$.}
\label{d}
\end{figure}

\begin{figure}
\includegraphics[width=8cm,clip=true]{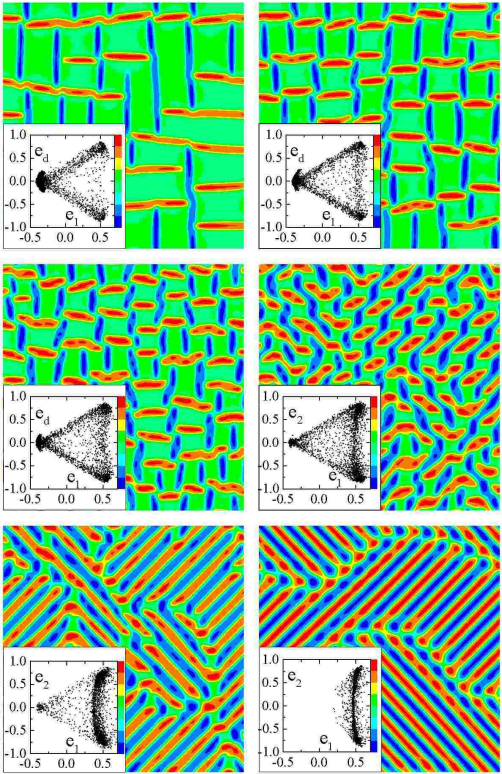}
\caption{(Color online) Morphologies for an anisotropic case in which the collapsed minima are located at $e_2=\pm e_2^C$, $e_3=0$. The
values of $\bar e_1$ for the plots are $-0.1$, 0, 0.1, 0.3, 0.4, and 0.5 respectively. 
Note that as long as there is some fraction of the system
in the isotropic elastic minimum (first three plots) the 
morphology of the collapsed regions is that of inter-crossed stripes along $x$
and $y$, while when the fraction of the system in the elastic configuration is negligible (last two plots) the morphology
changes to stripes along the diagonals.}
\label{rect}
\end{figure}

\begin{widetext}

\section{Appendix}

We give here the explicit expressions for the local free energy $f_0$ used in each of the simulations 
presented in the paper. For Figs. \ref{a1} and \ref{a2}, we use

\begin{equation}
f_0=\left [\frac{e_d^2}{2}+(e_1-1)^2\right ]\left [1+\tanh(3 e_1)\right ]+
\left [C_1\frac{e_d^2}{2}+(e_1+1)^2\right ]\left [1-\tanh(-3 e_1)\right ]
\end{equation}
with $C_1=1$ in Fig. \ref{a1}, and $C_1=0$ in Fig. \ref{a2}.
In Figs. \ref{b}, \ref{c}, \ref{d}, we use
\begin{equation}
f_0=\left [\frac{e_d^2}{2}+\frac{e_d^4}{4}+(e_1-C_3)^2\right ]\left [1+\tanh(C_2 e_1)\right ]+
\left [-\frac{e_d^2}{2}+\frac{e_d^4}{4}+(e_1+C_3)^2+\frac{1}{4}\right ]\left [1-\tanh(-C_2 e_1)\right ]
\end{equation}
with $C_2=2$ and $C_3=0.4$ in Fig. \ref{b}; $C_2=3.5$ and $C_3=0.2$ in Fig. \ref{c}; 
and $C_2=4$ and $C_3=0.1$ in Fig. \ref{d}. Finally, in Fig. \ref{rect}
we use
\begin{equation}
f_0=\left [\frac{e_2^2}{2}+\frac{e_2^4}{4}+(e_1-C_3)^2\right ]\left [1+\tanh(C_2 e_1)\right ]+
\left [-\frac{e_2^2}{2}+\frac{e_2^4}{4}+(e_1+C_3)^2+\frac{1}{4}\right ]\left [1-\tanh(-C_2 e_1)\right ]+e_3^2
\end{equation}
with $C_2=2$ and $C_3=0.4$.

\end{widetext}

\newpage 

\end{document}